\documentclass[conference]{IEEEtran}

\usepackage{epsfig,amsmath,amssymb,latexsym}
\usepackage{graphicx,epsfig,color,epsf,psfrag,hhline,amsmath,amssymb,textcomp}

\usepackage[ruled]{algorithm2e}

\def\_{\rule{.3em}{.15ex}}      

 \normalmarginpar
\newcommand {\mymarginpar}[1]{\marginpar{#1}}
\renewcommand {\marginpar}[1]{}

\def\_{\rule{.3em}{.15ex}}      

\newcommand{\ls}[1]
   {\dimen0=\fontdimen6\the\font
    \lineskip=#1\dimen0
    \advance\lineskip.5\fontdimen5\the\font
    \advance\lineskip-\dimen0
    \lineskiplimit=.9\lineskip
    \baselineskip=\lineskip
    \advance\baselineskip\dimen0
    \normallineskip\lineskip
    \normallineskiplimit\lineskiplimit
    \normalbaselineskip\baselineskip
    \ignorespaces
   }

\newcommand {\bearn}{\begin{eqnarray*}}
\newcommand {\eearn}{\end{eqnarray*}}
\newcommand {\barr}{\begin{array}}
\newcommand {\earr}{\end{array}}

\newcommand {\N}{{\cal N}}

\newtheorem{definition}{Definition}
\newtheorem{property}[definition]{Property}
\newtheorem{proposition}[definition]{Proposition}
\newtheorem{lemma}[definition]{Lemma}
\newtheorem{theorem}[definition]{Theorem}
\newtheorem{corollary}[definition]{Corollary}
\newtheorem{example}[definition]{Example}
\newtheorem{remark}[definition]{Remark}

\newcommand {\benum} {\begin{enumerate}}
\newcommand {\eenum} {\end{enumerate}}

\newcommand {\bdesc} {\begin{description}}
\newcommand {\edesc} {\end{description}}

\newcommand {\bfig}[2] {\begin{figure}
  \centering
  \includegraphics[width=#2]{#1}}
\newcommand {\brotatefig}[2] {\begin{figure}[htbp]
                        \centerline {
                         \epsfig{figure={#1},clip=,angle=-90,width={#2}}}}

\newcommand {\bfigfirst}[2] {\begin{figure}[h]
                        \centerline {
                        \setlength{\epsfxsize}{#2}
                        \epsffile{#1}}}
\newcommand {\efig}[2]{ \caption{#2}
                        \label{fig:#1}
                        \end{figure}
                        \mymarginpar{fig:#1}}
\newcommand {\erotatefig}[2]{ \caption{#2}
                        \label{fig:#1}
                        \end{figure}
                        \mymarginpar{fig:#1}}
\newcommand {\rfig}[1]{Figure \ref{fig:#1}}

\newcommand {\btab}[1]{
                       \begin{table}
                       \centering
                       \begin{tabular}{#1}}
\newcommand {\etab}[3] {
                       \end{tabular}
                       \caption[#3]{#2}
                       \label{tab:#1}
                       \end{table}
                       \mymarginpar{tab:#1}
                       \vspace{.1in}}

\newcommand {\btabular}[1]{\begin{center}
                       \begin{tabular}{#1}}
\newcommand {\etabular}{\end{tabular}
                       \end{center}}

\newcommand {\bdefin}[1]{\begin{definition}
                      \mymarginpar{def:#1}
                      \label{def:#1} }
\newcommand {\edefin}       {\end{definition}}
\newcommand {\rdef}[1]{Definition \ref{def:#1}}

\newcommand {\bpro}[1]{\begin{property}
                      \mymarginpar{pro:#1}
                      \label{pro:#1} }
\newcommand {\epro}   {\end{property}}

\newcommand {\bprop}[1]{\begin{proposition}
                      \mymarginpar{prop:#1}
                      \label{prop:#1} }
\newcommand {\eprop}       {\end{proposition}}
\newcommand {\rprop}[1]{Proposition \ref{prop:#1}}

\newcommand {\blem}[1]{\begin{lemma}
                      \mymarginpar{lem:#1}
                      \label{lem:#1} }
\newcommand {\elem}   {\end{lemma}}
\newcommand {\rlem}[1]{Lemma \ref{lem:#1}}

\newcommand {\bthe}[1]{\begin{theorem}
                      \mymarginpar{the:#1}
                      \label{the:#1} }
\newcommand {\ethe}   {\end{theorem}}
\newcommand {\rthe}[1]{Theorem \ref{the:#1}}

\newcommand {\bcor}[1]{\begin{corollary}
                      \mymarginpar{cor:#1}
                      \label{cor:#1} }
\newcommand {\ecor}   {\end{corollary}}

\newcommand {\bax}[1]{\begin{axiom}
                      \mymarginpar{ax:#1}
                      \label{ax:#1} }
\newcommand {\eax}       {\vspace{-.1in} \end{axiom}}

\newcommand {\bex}[2]{\vspace{.1in}
                      \begin{example}
                      \mymarginpar{ex:#1}
                       {\bf #2}
                      \label{ex:#1} }
\newcommand {\eex}       {\end{example} \vspace{.3cm} }

\newcommand {\brem}[1]{\begin{remark}
                      \mymarginpar{rem:#1}
                      \label{rem:#1} \em }
\newcommand {\erem}   {\end{remark}}

\newcommand {\beq}[1]{\mymarginpar{eq:#1}
                      \begin{equation}
                      \label{eq:#1} }

\newcommand {\beqno}[1]{\mymarginpar{eq:#1}
                      \begin{eqnarray}
                      \nonumber}

\newcommand {\eeq}       {\end{equation}}
\newcommand {\eeqno}       { && \end{eqnarray}}
\newcommand {\req}[1]{(\ref{eq:#1})}

\newcommand {\bear}[1]{\mymarginpar{eq:#1}
                       \begin{eqnarray}
                       \label{eq:#1} }

\newcommand {\bearno}[1]{\mymarginpar{eq:#1}
                       \begin{eqnarray}
                       \nonumber}

\newcommand {\eear}{\end{eqnarray}}
\newcommand {\eearno}{\end{eqnarray}}
\newcommand {\bsel}{\left \{ \begin{array}{cl}}
\newcommand {\esel}{\end{array} \right.}

\newcommand {\bmat}[1]{\left [ \begin{array}{#1}}
\newcommand {\emat}{\end{array} \right ]}
\newcommand {\bsec}[2]{\mymarginpar{sec:#2}
                       \section{#1}
                       \label{sec:#2} }

\newcommand {\rsec}[1]{Section \ref{sec:#1}}

\newcommand {\bsubsec}[2]{\mymarginpar{sec:#2}
                       \subsection{#1}
                       \label{sec:#2} }

\def\R{I\kern-0.30em R}
\def\N{I\kern-0.30em N}
\def\P{I\kern-0.30em P}

\def\pr{{\bf\sf P}}

\newcommand\sg{(G(V_g,E_g), p(\cdot,\cdot))}
\newcommand\RC{C}

\newcommand\qq{q}

\begin{document}

\title{A Probabilistic Framework for Structural Analysis in Directed Networks}


\author{Cheng-Shang Chang, Duan-Shin Lee, Li-Heng Liou, Sheng-Min Lu, and Mu-Huan Wu\\
Institute of Communications Engineering,
National Tsing Hua University \\
Hsinchu 30013, Taiwan, R.O.C. \\
Email:  cschang@ee.nthu.edu.tw; lds@cs.nthu.edu.tw; dacapo1142@gmail.com;\\
 s103064515@m103.nthu.edu.tw; u9661106@oz.nthu.edu.tw;\\
}



\maketitle

\begin{abstract}

In our recent works, we developed a probabilistic framework for structural analysis in {\em undirected networks}.
The key idea of  that framework is to sample a network by a {\em symmetric} bivariate distribution and then use that bivariate distribution to formerly define various notions, including  centrality, relative centrality, community, and modularity.
The main objective of this paper is to extend the probabilistic framework  to {\em directed} networks,
where the sampling bivariate distributions could be {\em asymmetric}.
Our main finding is that we can relax the assumption from  {\em symmetric} bivariate distributions to bivariate distributions that have the same marginal distributions. By using such a weaker assumption, we show that various  notions for structural analysis  in directed networks can also be defined in the same manner as before. However, since the bivariate distribution could be {\em asymmetric}, the  community detection algorithms proposed in our previous work  cannot be directly applied. For this,  we show that one can construct another sampled graph with a {\em symmetric} bivariate distribution so that for any partition of the network,  the modularity index  remains the same as that of the original sampled graph. Based on this, we propose a hierarchical agglomerative  algorithm that returns a partition of communities when the algorithm converges.


\end{abstract}


{\bf keywords:} centrality, community, modularity, PageRank


\bsec{Introduction}{introduction}

As the advent of on-line social networks, structural analysis of networks has been a very hot research topic. There are various notions that are widely used
 for structural analysis of networks, including centrality, relative centrality, similarity, community, modularity, and homophily (see e.g., the book by Newman \cite{Newman2010}).
In order to make these notions more mathematically precise, we developed in \cite{Chang:11:AGP,chang2013relative} a probabilistic framework for structural analysis of {\em undirected} networks. The key idea of the framework is to ``sample'' a network to generate a bivariate distribution $p(v,w)$ that specifies the probability that a pair of two nodes $v$ and $w$ are selected from a sample. The bivariate distribution  $p(v,w)$ can be viewed as a normalized {\em similarity} measure \cite{LK03} between the two nodes $v$ and $w$.  A graph $G$ associated with a bivariate distribution $p(\cdot,\cdot)$ is then called a {\em sampled} graph.

In \cite{Chang:11:AGP,chang2013relative}, the bivariate distribution is assumed to be {\em symmetric}. Under this assumption, the two marginal distributions of the bivariate distribution, denoted by $p_V(\cdot)$ and $p_W(\cdot)$, are the same and they represent the probability that a particular node is selected in the sampled graph. As such, the marginal distribution $p_V (v)$ can be used for defining the {\em centrality} of a node $v$ as it represents the probability that node $v$ is selected.
 The relative centrality of a set of nodes $S_1$ with respect to another set of nodes $S_2$ is then defined as the {\em conditional} probability that one node of the selected pair of two nodes is in the set $S_1$ given that the other node is in the set $S_2$.
Based on the probabilistic definitions of centrality and relative centrality in the framework, the {\em community strength} for a set of nodes $S$ is defined as the difference between its relative centrality with respect to itself and its centrality. Moreover,  a set of nodes with a {\em nonnegative}  community strength  is called a {\em community}.
In the probabilistic framework, the {\em modularity} for a partition of a sampled graph is defined as the average community strength of the community. As such, a high modularity for a partition of a graph implies that there are  communities with strong community strengths.  It was further shown in \cite{chang2013relative} that the Newman modularity in \cite{NG04} and the stability in \cite{Lambiotte2010,Delvenne2010} are special cases of the modularity for certain sampled graphs.


The main objective of this paper is to extend the probabilistic framework in \cite{Chang:11:AGP,chang2013relative} to {\em directed} networks,
where the sampling bivariate distributions could be {\em asymmetric}.
Our main finding is that we can relax the assumption from  {\em symmetric} bivariate distributions to bivariate distributions that have the same marginal distributions. By using such a weaker assumption, we show that the notions of centrality, relative centrality, community and modularity can be defined in the same manner as before. Moreover, the equivalent characterizations of a community still hold. Since the bivariate distribution could be {\em asymmetric}, the agglomerative community detection algorithms in \cite{Chang:11:AGP,chang2013relative} cannot be directly applied.
For this,  we show that one can construct another sampled graph with a {\em symmetric} bivariate distribution so that for any partition of the network,  the modularity index  remains the same as that of the original sampled graph. Based on this, we propose a hierarchical agglomerative  algorithm that returns a partition of communities when the algorithm converges.

 In this paper, we also address two methods for sampling a directed network with a bivariate distribution that has the same marginal distributions : (i) PageRank and (ii) random walks with self loops and backward jumps.
      Experiments show that sampling by a random walk with self loops and backward jumps performs better than that by PageRank for community detection. This might be due to the fact that PageRank adds weak links in a network and that changes the topology of the network and thus affects the results of community detection.

\bsec{Sampling networks by bivariate distributions with the same marginal distributions}{sampling}

In \cite{chang2013relative}, a probabilistic framework for network analysis for undirected networks was proposed.
The main idea in that framework is to characterize a network by a  {\em sampled graph}.
Specifically, suppose a network is modelled by a graph $G(V_g,E_g)$, where $V_g$ denotes the set of vertices (nodes) in the graph and $E_g$ denotes the set of edges (links) in the graph.
  Let $n=|V_g|$ be the number of vertices in the graph and index the $n$ vertices from $1,2,\ldots, n$. Also, let $A=(a_{ij})$ be the $n \times n$ adjacency matrix of the graph, i.e.,
\bearn
a_{ij}
=  \left\{\begin{array}{ll}
                 1, & \mbox{if there is an edge from  vertex {\em i} to vertex {\em j}}, \\
                 0, & $otherwise$.
                \end{array} \right.
\eearn

A sampling bivariate distribution $p(\cdot,\cdot)$ for a graph $G$ is the bivariate distribution that is used for  {\em sampling} a network by randomly selecting an ordered pair of two nodes $(V,W)$, i.e.,
\beq{frame1111}
\pr(V=v,W=w)=p(v,w).
\eeq
Let $p_V(v)$ (resp. $p_W(w)$) be the marginal distribution of the random variable $V$ (resp. $W$), i.e.,
\beq{frame2222}
p_V(v)=\pr(V=v)= \sum_{w=1}^n p(v,w),
\eeq
and
\beq{frame3333}
p_W(w)= \pr(W=w)=\sum_{v=1}^n p(v,w).
\eeq

\bdefin{sampled}{\bf (Sampled graph)} A  graph $G(V_g,E_g)$ that is sampled by randomly selecting an ordered pair of two nodes $(V,W)$ according to a specific bivariate distribution $p(\cdot,\cdot)$  in \req{frame1111} is called a {\em sampled graph} and it is denoted by the two tuple $\sg$.
\edefin

For a given graph $G(V_g, E_g)$, there are many methods to generate sampled graphs by specifying the needed bivariate distributions. In \cite{chang2013relative}, the bivariate distributions are all assumed to be {\em symmetric} and that limits its applicability to {\em undirected} networks.
One of the main objectives of this paper
 is to relax the {\em symmetric} assumption for the bivariate distribution so that the framework can be applied to {\em directed} networks. The key idea of doing this is to assume that the bivariate distribution has the same marginal distributions, i.e.,
 \beq{iden1111}
 p_V(v)=p_W(v), \quad \mbox{for all}\;v.
 \eeq
 Note that a symmetric bivariate distribution has the same marginal distributions and thus the assumption in \req{iden1111} is much more general.



\bsubsec{PageRank}{page}

One approach for sampling a network with a bivariate distribution that has the same marginal distributions  is to sample a network by an {\em  ergodic Markov chain}. From the Markov chain theory (see e.g., \cite{nelson1995probability}), it is well-known that an ergodic Markov chain converges to its steady state in the long run. Hence, the joint distribution of two successive steps of a {\em stationary and ergodic} Markov chain can be used as the needed bivariate distribution.
Specifically, suppose that a network $G(V_g, E_g)$ is sampled by a stationary and ergodic Markov chain $\{X(t), t \ge 0\}$ with the state space
$\{1,2, \ldots, n\}$ being the $n$ nodes in $V_g$. Let $P=(p_{ij})$ be the $n \times n$ transition probability matrix  and ${\bf \pi}=(\pi_1, \pi_2, \ldots, \pi_n)$ be the steady state probability vector of the stationary and ergodic Markov chain. Then we can choose the bivariate distribution
\bear{map1111}
&&\pr(V=v,W=w)=p(v,w)\nonumber \\
&&=\pr (X(t)=v, X(t+1)=w).
\eear
As the Markov chain is stationary, we have
\beq{map2222}
\pr (X(t)=v)=\pr (X(t+1)=w)=p_V(v)=p_W(v).
\eeq

It is well-known that a random walk on the graph induces a Markov chain with the state transition probability matrix
$P=(p_{ij})$ with
\beq{rwalk1111}
p_{ij}={{a_{ij}} \over {k^{out}_i}},
\eeq
where
\beq{rwalk2222}
k^{out}_i=\sum_{j=1}^n a_{ij},
\eeq
is the number of outgoing edges from vertex $i$.
In particular, if the graph is an undirected graph, i.e., $a_{ij}=a_{ji}$, then the induced Markov chain is reversible and the steady state probability  of state $i$, i.e., $\pi_i$, is $k_i/2m$, where
$m={1 \over 2}\sum_{i=1}^n \sum_{j=1}^n a_{ij}$ is the total number of edges of the undirected graph.

One problem for sampling
 a {\em directed} network by a simple random walk is that the induced Markov chain may not be ergodic even
when the network itself is weakly connected.
One genuine solution for this is to allow random jumps from states to states in a random walk. PageRank \cite{PageRank}, proposed by Google, is one such example that has been successfully used for ranking web pages. The key idea behind PageRank is to model the behavior of a web surfer by a random walk (the random surfer model) and  then use that to compute the steady state probability for a web surfer to visit a specific web page.
Specifically, suppose that there are $n$ web pages and a web surfer uniformly selects a web page  with probability $1/n$. Once he/she is on a web page,  he/she continues web surfing with probability $\lambda$. This is done by selecting {\em uniformly} one of the hyperlinks in that web page.  On the other hand, with probability $1-\lambda$ he/she starts a new web page {\em uniformly} among all the $n$ web pages.
 The transition probability from state $i$ to state $j$ for the induced Markov chain is then
\beq{Page0000}
p_{ij}= (1-\lambda){1 \over n} + \lambda {a_{ij} \over {k_i^{out}}},
\eeq
where $a_{ij}=1$ if there is a hyperlink pointing from the $i^{th}$ web page to the $j^{th}$ web page and  $k_i^{out}=\sum_{j=1}^n a_{ij}$ is the total number of hyperlinks on the $i^{th}$ web page.
Let $\pi_i$ be steady probability of visiting the $i^{th}$ web page by the web surfer. It then follows that
\beq{Page1111}
\pi_i =(1-\lambda){1 \over n} + \lambda \sum_{j=1}^n {a_{ji} \over {k_j^{out}}} \pi_j .
\eeq
PageRank then uses $\pi_i$ as the centrality of the $i^{th}$ web page and rank web pages by their centralities. Unlike the random walk on an undirected graph, the steady state probabilities in \req{Page1111} cannot be explicitly solved and it requires a lot of computation to solve the system of linear equations.

The sampled graph $\sg$ by using PageRank then has the following bivariate distribution
\beq{Page0088}
p(v,w)=\pi_v p_{vw},
\eeq
where $p_{vw}$ is defined in \req{Page0000} and $\pi_v$ is the solution of \req{Page1111}.

\bsubsec{Random walks with self loops and backward jumps}{selfloop}

Another way to look at  the Markov chain induced by PageRank in \req{Page0000} is that it is in fact a random walk on a different graph with the adjacency matrix $\tilde A$ that is constructed from the original graph with additional edge weights, i.e.,
\beq{PageRank1111}
\tilde A= (1 -\lambda) {1 \over n} {\bf 1}+ \lambda D^{-1} A,
\eeq
where ${\bf 1}$ is an $n \times n$ matrix with all its elements being 1 and
$D=(d_{ij})$ is the diagonal matrix with  $d_{ii}=k_i^{out}$ for all $i=1,2,\ldots, n$.

In view of \req{PageRank1111}, another solution for the ergodic problem is to consider a random walk on the graph with the adjacency matrix
\beq{PageRank2222}
\hat A= \lambda_0 {\bf I} + \lambda_1 A+ \lambda_2 A^T,
\eeq
where $\bf I$ is the $n \times n$ identity matrix and $A^T$ is the transpose matrix of $A$.
The three parameters  $\lambda_0, \lambda_1$, $\lambda_2$ are positive and
 $$\lambda_0+\lambda_1+ \lambda_2 =1.$$
A random walk on the graph with the adjacency matrix $\hat A$  induces an ergodic Markov chain if the original graph is weakly connected. Also, with the additional edges from the identity matrix and the transpose matrix,  such a random walk can be viewed as a random walk on the original graph with self loops and backward jumps.

\bsec{The framework for directed networks}{framework}

\bsubsec{Centrality and  relative centrality}{relcen}

Centrality \cite{freeman1977set,freeman1979centrality,Newman2010} is usually used as a measure for ranking the importance of a set of nodes in a (social) network. Under the assumption in \req{iden1111}, such a concept can
be directly mapped to the probability that a node is selected as in \cite{chang2013relative}.


\bdefin{centralitym}
{\bf (Centrality)} For a sampled graph $\sg$ with the bivariate distribution $p(\cdot,\cdot)$ that has the same marginal distributions in \req{iden1111},
the {\em centrality} of a set of nodes $S$, denoted by $C(S)$, is defined as the probability that a node in $S$ is selected, i.e.,
\beq{central1111m}
C(S)=\pr(V \in S)=\pr (W \in S) .
\eeq
\edefin

As a generalization of  centrality,
relative centrality in \cite{chang2013relative} is a (probability) measure that measures how important a set of nodes in a network is with respect to another set of nodes.

\bdefin{relativem} {\bf (Relative centrality)}
For a sampled graph $\sg$ with the bivariate distribution $p(\cdot,\cdot)$ that has the same marginal distributions in \req{iden1111},
the {\em relative centrality} of a set of nodes $S_1$ with respect to another set of nodes $S_2$, denoted by $\RC(S_1|S_2)$, is defined as the conditional probability
that  the randomly selected node $W$ is inside $S_1$ given that the random selected node $V$ is inside $S_2$, i.e.,
\beq{relative0000m}
\RC(S_1| S_2)=\pr(W \in S_1 |V \in S_2)
\eeq
\edefin

We note that if we choose $S_2=V_g$,
then the relative centrality of a set of nodes $S_1$ with respect to $V_g$ is simply the {\em centrality} of the set of nodes $S_1$.

\bex{pagerel}{\bf (Relative PageRank)}
PageRank described in \rsec{page} has been commonly used for ranking the importance of nodes in a directed network. Here we can use \rdef{relativem} to define relative PageRank that can be used for ranking the relative importance of a set of nodes to another set of nodes in a directed network. Specifically,
let $\pi$ be the PageRank for node $i$ in \req{Page1111} and $p_{i,j}$ be the transition probability from state $i$ to state $j$ for the induced Markov chain in \req{Page0000}. Then the relative PageRank of a set $S_1$ with respect to another set $S_2$ is
\bear{relative1111m}
&&\RC(S_1| S_2)=\pr(W \in S_1 |V \in S_2)\nonumber \\
&&={{\pr (W \in S_1 , V \in S_2)} \over {\pr (V \in S_2)}}
={{\sum_{i \in S_2}\sum_{j \in S_1}\pi_i p_{ij}} \over {\sum_{i \in S_2}\pi_i}}.
\eear
\eex

Analogous to the relative centrality in \cite{chang2013relative}, there are also several properties of relative centrality in
\rdef{relativem}. However, the reciprocity property in \rprop{relprom}(iv) is much weaker than that in \cite{chang2013relative}.
The proof of \rprop{relprom} is given in Appendix A.

\bprop{relprom} For a sampled graph $\sg$ with the bivariate distribution $p(\cdot,\cdot)$  that has the same marginal distributions in \req{iden1111},
the following properties for the relative centrality defined in \rdef{relativem} hold.

\noindent (i) $0 \le \RC(S_1| S_2) \le 1$ and $0 \le C(S_1) \le 1$. Moreover,
$\RC(V_g| S_2)=1$ and $C(V_g)=1$.

\noindent (ii) (Additivity) If $S_1$ and $S_2$ are two disjoint sets., i.e., $S_1 \cap S_2$ is an empty set, then
for an arbitrary set $S_3$,
\bear{merge1123}
\RC (S_1 \cup S_2 | S_3)
= \RC (S_1 | S_3) + \RC (S_2 | S_3).
\eear
In particular, when $S_3= \{1,2, \ldots, n\}$, we have
\beq{merge1124}
C(S_1 \cup S_2) =C(S_1) + C(S_2).
\eeq

\noindent (iii) (Monotonicity) If $S_1$ is a subset of $S_1^\prime$, i.e., $S_1 \subset S_1^\prime$, then
$\RC(S_1| S_2)\le \RC(S_1^\prime| S_2)$ and $C(S_1)\le C(S_1^\prime)$.

\noindent (iv)
(Reciprocity) Let $S^c =V_g \backslash S$ be the set of nodes that are {\em not} in $S$.
$$C(S) \RC(S^c| S)= C(S^c) \RC(S| S^c).$$
\eprop

\bsubsec{Community strength and communities}{comdefinition}

The notions of community strength and modularity in \cite{chang2013relative} generalizes the original Newman's definition \cite{Newman04} and unifies various other generalizations, including the stability in \cite{Lambiotte2010,Delvenne2010}.
In this section,  we further extend these notions to directed networks.

\bdefin{community} {\bf (Community strength and communities)} For a sample graph $\sg$ with a  bivariate distribution $p(\cdot,\cdot)$ that has the same marginal distributions in \req{iden1111},
the {\em community strength} of
a subset set of nodes $S \subset V_g$, denoted by $Str(S)$,  is defined as the difference of the relative centrality of $S$ with respect to itself and its centrality, i.e.,
\beq{community1111}
Str(S)=\RC(S|S)-C(S).
\eeq
In particular, if a subset of nodes $S \subset V_g$ has a nonnegative community strength, i.e., $Str(S) \ge 0$, then it  is  called
a {\em community}.
\edefin

In the following theorem, we show various equivalent statements for a set of nodes to be a community.
The proof of \rthe{communeq} is given in Appendix B.

\bthe{communeq}
Consider a sample graph $\sg$ with a  bivariate distribution $p(\cdot,\cdot)$ that has the same marginal distributions in \req{iden1111},
 and a set $S$ with $0 < C(S)<1$. Let $S^c = V_g \backslash S$ be the set of nodes that are not in $S$.
The following statements are equivalent.
\begin{description}
\item[(i)] The set $S$ is a community, i.e., $Str(S)=\RC(S|S)-C(S)\ge 0$.
\item[(ii)]  The relative centrality of $S$ with respect to $S$ is not less than the relative centrality of $S$ with respect to $S^c$, i.e., $\RC(S|S) \ge \RC(S|S^c)$.
\item[(iii)] The relative centrality of $S^c$ with respect to $S$ is not greater than the centrality of $S^c$, i.e., $\RC(S^c|S) \le C(S^c)$.
    \item[(iv)]   The relative centrality of $S$ with respect to $S^c$ is not greater than the centrality of $S$, i.e., $\RC(S|S^c) \le C(S)$.
    \item[(v)] The set $S^c$ is a community, i.e., $Str(S^c)=\RC(S^c|S^c)-C(S^c)\ge 0$.
    \item[(vi)] The relative centrality of $S^c$ with respect to $S^c$ is not less than the relative centrality of $S^c$ with respect to $S$, i.e., $\RC(S^c|S^c) \ge \RC(S^c|S)$.
\end{description}
\ethe

\bsubsec{Modularity and community detection}{comdet}

As in \cite{chang2013relative}, we define the modularity index for a partition of a network as the average community strength of a randomly selected node in \rdef{index}.

\bdefin{index}{\bf (Modularity)}
Consider a sampled graph $\sg$ with a  bivariate distribution $p(\cdot,\cdot)$ that has the same marginal distributions in \req{iden1111}.
Let ${\cal P}=\{S_c, c=1,2, \ldots, C\}$, be a partition of $\{1,2,\ldots, n\}$, i.e.,
$S_c \cap S_{c^\prime}$ is an empty set for $c \ne c^\prime$ and $\cup_{c=1}^C S_c=\{1,2,\ldots, n\}$.
The  modularity index $Q({\cal P})$ with respect to the partition $S_c$, $c=1,2, \ldots, C$, is
defined as the weighted average of the community strength of each subset with the weight being  the centrality of each subset, i.e.,
\beq{index1111}
Q({\cal P})=\sum_{c=1}^C C(S_c)\cdot Str(S_c).
\eeq
\edefin

We note the modularity index in \req{index1111} can also be written as follows:

$$Q({\cal P})=\sum_{c=1}^C \pr (V \in S_c, W \in S_c) -\pr (V \in S_c) \pr (W \in S_c)$$
\bear{index2222}
=\sum_{c=1}^C \sum_{v \in S_c} \sum_{w \in S_c} (p(v,w)-p_V(v)p_W(w)).
\eear

As the modularity index for a partition of a network is the average community strength of a randomly selected node,
 a good partition of a network should have a large modularity index. In view of this, one can then tackle the community detection problem by looking for algorithms that yield large values of the modularity index.
  For sampled graphs with {\em symmetric} bivariate distributions, there are already various community detection algorithms in \cite{Chang:11:AGP,chang2013relative} that find local maxima of the modularity index.
 However, they cannot be directly applied as the bivariate distributions for sampling directed networks could be {\em asymmetric}.
 For this,  we show in the following lemma  that one can construct another sampled graph with a {\em symmetric} bivariate distribution so that for any partition of the network,  the modularity index  remains the same as that of the original sampled graph.  The proof of \rlem{transform} is given in Appendix C.

\blem{transform}
Consider a sampled graph $\sg$ with a  bivariate distribution $p(\cdot,\cdot)$ that has the same marginal distributions in \req{iden1111}.
Construct the sampled graph $(G(V_g, E_g), \tilde p(\cdot,\cdot))$ with the symmetric bivariate distribution
\beq{tran1111}
 \tilde p(v,w)={{p(v,w)+p(w,v)} \over 2}.
 \eeq
Let $Q({\cal P})$ (resp. $\tilde Q({\cal P})$) be the modularity index for the partition ${\cal P}=\{S_c, c=1,2, \ldots, C\}$ of the sampled graph $\sg$ (resp. the sampled graph $(G(V_g, E_g), \tilde p(\cdot,\cdot))$). Then
\beq{tran2222}
\tilde Q({\cal P})=Q({\cal P}).
\eeq
\elem

As $\tilde Q({\cal P})=Q({\cal P})$, one can then use the community detection algorithms for the  sampled graph $(G(V_g, E_g), \tilde p(\cdot,\cdot))$ with the symmetric bivariate distribution to solve the community detection problem for the original sampled graph $\sg$.
Analogous to the hierarchical agglomerative algorithms in  \cite{Newman04,blondel2008fast},
 in the following we propose  a hierarchical agglomerative algorithm for community detection in directed networks.
 The idea behind this algorithm is {\em modularity maximization}. For this, we define the correlation measure between two nodes $v$ and $w$
  as follows:
\bear{random4405}
&&\qq(v,w) = \tilde p(v,w)-\tilde p_V(v) \tilde p_W(w) \nonumber \\
&&=\frac{p(v,w)-p_V(v)p_W(w)+p(w,v)-p_V(w)p_W(v)}{2}. \nonumber \\
\eear
For any two sets $S_1$ and $S_2$, define the correlation measure between these two sets as
\beq{random4406}
\qq(S_1, S_2)=\sum_{v \in S_1}\sum_{w \in S_2} \qq(v,w).
\eeq
Also, define the {\em average} correlation measure between two sets $S_1$ and $S_2$ as
\beq{avgcoh1111}
\bar \qq (S_1, S_2) ={1 \over {|S_1|\cdot |S_2|}}\qq(S_1, S_2).
\eeq
With this correlation measure,
 we have from \rlem{transform}, \req{index2222} and \req{random4406} that
the modularity index for the partition ${\cal P}=\{S_c, c=1,2, \ldots, C\}$ is
\beq{random4407}
Q({\cal P})=\tilde Q ({\cal P})=\sum_{c=1}^C \qq(S_c,S_c),
\eeq
Moreover, a set $S$ is a community if and only if $\qq(S, S) \ge 0$.

\noindent{\bf Algorithm 1: a hierarchical agglomerative  algorithm for community detection in a directed network}


\noindent {\bf (P0)} Input a sampled graph $\sg$ with a  bivariate distribution $p(\cdot,\cdot)$ that has the same marginal distributions in \req{iden1111}.

\noindent {\bf (P1)} Initially, there are $n$ sets, indexed from 1 to $n$, with each set containing exactly one node. Specifically, let $S_i$ be the set of nodes in set $i$. Then $S_i=\{i\}$, $i=1,2,\ldots, n$.

\noindent {\bf (P2)}  For all $i, j=1,2, \ldots, n$, compute the correlation measures
$\qq(S_i, S_j)=\qq(\{i\},\{j\})$ from \req{random4405}.

\noindent {\bf  (P3)} If there is only one set left or there do not exist nonnegative correlation measures between two distinct sets, i.e., $\qq(S_i, S_j)<0$ for all $i \ne j$, then the algorithm outputs the current sets.

\noindent {\bf (P4)} Find two  sets that have a nonnegative correlation measure. Merge these two sets
into a new set. Suppose that set $i$ and set $j$ are grouped into a new set $k$.
Then $S_k =S_i \cup S_j$ and update
\beq{random4400}
\qq(S_k, S_k)= \qq(S_i, S_i)+2 \qq (S_i, S_j) + \qq (S_j, S_j).
\eeq
 Moreover, for all $\ell \ne k$, update
 \beq{random4455}
\qq (S_k, S_\ell) =\qq(S_\ell, S_k)=\qq(S_i, S_\ell)+\qq(S_j, S_\ell).
\eeq

\noindent {\bf (P5)} Repeat from (P3).


The hierarchical agglomerative  algorithm in Algorithm 1 has the following properties.

\bthe{community}
\begin{description}
\item[(i)] For the hierarchical agglomerative  algorithm in Algorithm 1,
the modularity index is non-decreasing in every iteration and thus converges to a local optimum.
\item[(ii)] When the algorithm converges, every set returned by the hierarchical agglomerative algorithm  is indeed
a {\em community}.
\item[(iii)] If, furthermore, we use the {\em greedy} selection that
selects the two sets with the {\em largest} average correlation measure to merge in (P4) of Algorithm 1, then the average correlation  measure of the two selected sets in each merge operation is non-increasing.
\end{description}
\ethe

The proof of \rthe{community} is given in Appendix D. For (i) and (ii) of \rthe{community}, it is not necessary to specify how we select a pair of two sets with a nonnegative correlation.
One advantage of using the greedy selection in (iii) of \rthe{community} is the {\em monotonicity} property for the dendrogram produced by a greedy hierarchical agglomerative algorithm (see \cite{theodoridispattern}, Chapter 13.2.3).
With such a monotonicity property, there is no {\em crossover} in the produced dendrogram.

\bsec{Experimental results}{experiments}

In this section, we compare the sampling methods by PageRank in \rsec{page} and random walks with self loops and backward jumps in \rsec{selfloop} for community detection. We conduct various experiments based on the stochastic block model with two blocks.
The stochastic block model, as a generalization of the Erdos-Renyi random graph, is a commonly used method for generating random graphs that can be used for benchmarking community detection algorithms. In a stochastic block model with two blocks (communities), the total number of nodes in the random graph are evenly distributed to these two blocks. The probability that there is an edge between two nodes within the same block is $p_{in}$ and the probability that there is an edge between two nodes in two different blocks is $p_{out}$. These edges are generated independently. Let $c_{in}=n p_{in}$ and  $c_{out}=n p_{out}$.

\begin{figure}[h]
\centering
\includegraphics[width=0.4\textwidth]{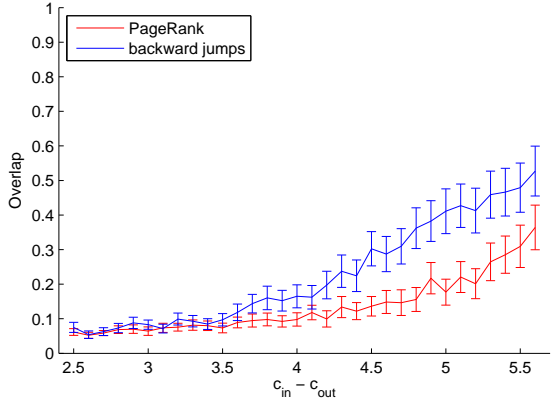}
\caption{Community detection of the stochastic block model by using PageRank in \req{PageRank1111}
and a random walk with self loops and backward jumps in \req{PageRank2222}.}
\label{fig:sbm}
\end{figure}

In our experiments, the number of nodes $n$ in the stochastic block model is 200 with 100 nodes in each of these two blocks. The average degree of a node is set to be 3. The values of $c_{in}-c_{out}$ of these graphs are in the range from 2.5 to 5.9 with a common step of 0.1. We generate 100 graphs for each $c_{in}-c_{out}$. Isolated vertices are removed. Thus, the exact numbers of vertices used in this experiment might be slightly less than 200.
For PageRank, the parameter $\lambda$ is chosen to be 0.9.
For the random walk with self loops and backward jumps, the three parameters are $\lambda_0=0.05$, $\lambda_1=0.85$ and $\lambda_2=0.1$.
We run the greedy hierarchical agglomerative  algorithm in Algorithm 1 until there are only two sets (even when  there do not exist nonnegative correlation measures between two distinct sets). We then evaluate the overlap with the true labeling. In \rfig{sbm}, we show the experimental results, where  each point is averaged over 100 random graphs from the stochastic block model. The error bars are the $95\%$ confidence intervals.
From \rfig{sbm}, one can see that the performance of random walks with self loops and backward jumps  is better than that of PageRank.  One reason for this is that PageRank uniformly adds an edge (with a small weight) between any two nodes and these added edges change the network topology. On the other hand, mapping by a random walk with backward jumps in \req{PageRank2222} does not change the network topology when it is viewed as an undirected network.

\bsec{Conclusion}{conclusion}


  In this paper we extended our previous work in \cite{Chang:11:AGP,chang2013relative} to {\em directed networks}. Our approach is to introduce bivariate distributions that have the same marginal distributions.  By doing so, we were able to extend the notions of centrality, relative centrality,  community strength, community and modularity to {\em directed networks}. For community detection, we propose a hierarchical agglomerative algorithm that guarantees every set returned from the algorithm is a community. We also tested the algorithm
  by using PageRank and random walks with self loops and backward jumps. The experimental results show that sampling by  random walks with self loops and backward jumps perform better than sampling by PageRank for community detection.

%
\bibliographystyle{IEEEtran}

\section*{Appendices}

\subsection*{Appendix A}

In this section , we prove \rprop{relprom}.
Since the relative centrality is a conditional probability and the centrality is a probability, the properties in (i),(ii) and (iii) follow trivially from the property of probability measures.

(vi)
 From \req{relative0000m} and \req{central1111m}, it follows that
\bear{relative3355m}
&&C(S) \RC(S^c| S) =\pr ( W \in S) \pr (W \in S^c | V \in S) \nonumber \\
&&= \pr (V \in S) \pr (W \in S^c | V \in S)\nonumber \\
&&= \pr (V \in S, W \in S^c).
\eear
Similarly, we also have
\bear{relative3366m}
&& C(S^c) \RC(S| S^c)=\pr (V \in S^c, W \in S).
 \eear
 Thus, it suffices to show that
 \beq{total0000}
 \pr (V \in S, W \in S^c)=\pr (V \in S^c, W \in S).
 \eeq
Note  that
\beq{total1111a}
\pr (V \in S)=\pr (V \in S, W \in S)+\pr (V \in S, W \in S^c),
\eeq
and
\beq{total1111b}
\pr (W \in S)=\pr (V \in S, W \in S)+\pr (V \in S^c, W \in S).
\eeq
Since the bivariate distribution has the same marginal distributions, we have $\pr(V \in S)=\pr(W \in S)$.
In conjunction with \req{total1111a} and \req{total1111b}, we prove \req{total0000}.

\subsection*{Appendix B}

In this section, we prove \rthe{communeq}.
We first prove that the first four statements are equivalent by showing (i)$\Rightarrow$ (ii)$\Rightarrow$ (iii)$\Rightarrow$ (iv)$\Rightarrow$ (i).

(i) $\Rightarrow$ (ii): Note from \rprop{relprom} (i) and (ii) that $\RC(S|S)+\RC(S^c|S)=\RC(V_g|S)=1$ and $C(S)+C(S^c)=C(V_g)=1$.
It then follows from the reciprocal property in \rprop{relprom}(iv) that
\bearn
&&C(S^c) (\RC(S|S) - \RC(S|S^c) )\\
&&=C(S^c)\RC(S|S) - C(S^c) \RC (S|S^c) \\
&&=(1-C(S))\RC(S|S) - C(S) \RC(S^c |S)\\
&&=(1-C(S))\RC(S|S) - C(S) (1- \RC(S |S))\\
&&=\RC(S|S)-C(S)= Str(S) \ge 0.
\eearn
As we assume that $0 < C(S) <1$, we also have $0 < C(S^c) < 1$. Thus,
$$\RC(S|S) - \RC(S|S^c) \ge 0.$$

(ii) $\Rightarrow$ (iii): Since we assume that $\RC(S|S) \ge \RC(S|S^c)$,
we have from $\RC(S|S)+\RC(S^c|S)=\RC(V_g|S)=1$  that
$$1= \RC (S|S) + \RC (S^c |S) \ge \RC(S|S^c)+\RC(S^c |S).$$
Multiplying both sides by $C(S^c)$ yields
$$C(S^c) \ge C(S^c)\RC(S|S^c) + C(S^c) \RC(S^c|S).$$
From the reciprocal property in \rprop{relprom}(iv) and $C(S)+C(S^c)=C(V_g)=1$, it follows that
\bearn
C(S^c) &\ge& C(S)\RC(S^c|S) + C(S^c) \RC(S^c|S) \\
&=&(C(S)+C(S^c))\RC(S^c|S) \\
&=&\RC(S^c|S).
\eearn

(iii) $\Rightarrow$ (iv): Note from the reciprocal property in \rprop{relprom}(iv) that
\beq{eqiv1111}
C(S)\RC (S^c| S) = C(S^c) \RC (S| S^c).
\eeq
It then follows from $\RC(S^c|S) \le C(S^c)$ that $\RC(S|S^c) \le C(S)$.

(iv) $\Rightarrow$ (i): Since we assume that $\RC(S|S^c) \le C(S)$, it follows from \req{eqiv1111}
that $\RC(S^c|S) \le C(S^c)$. In conjunction with $\RC(S|S)+\RC(S^c|S)=\RC(V_g|S)=1$ and $C(S)+C(S^c)=C(V_g)=1$,
we have
$$\RC(S|S)-C(S)=C(S^c)-\RC(S^c|S) \ge 0.$$

Now we show that and (iv) and (v) are equivalent.
Since $\RC(S|S^c)+\RC(S^c|S^c)=\RC(V_g|S^c)=1$ and $C(S)+C(S^c)=C(V_g)=1$, we have
$$\RC(S|S^c)-C(S)=C(S^c)-\RC(S^c|S^c)=-Str(S^c).$$
Thus, $\RC(S|S^c)\le C(S)$ if and only if $Str(S^c) \ge 0$.

Replacing $S$ by $S^c$, we see that (v) and (vi) are also equivalent because (i) and (ii) are equivalent.


\subsection*{Appendix C}

In this section, we prove \rlem{transform}.

Since $p_V(v)=p_W(v)$ for all $v$, it then follows from \req{index2222} that
\bear{index3333}
Q({\cal P})&=&\sum_{c=1}^C \sum_{v \in S_c} \sum_{w \in S_c} (p(v,w)-p_V(v)p_V(w)) \label{eq:index3333a} \\
&=&\sum_{c=1}^C \sum_{v \in S_c} \sum_{w \in S_c} (p(w,v)-p_V(w)p_V(v)) \label{eq:index3333b}.
\eear
Adding \req{index3333a} and \req{index3333b} yields
\beq{index4444}
2Q({\cal P})=\sum_{c=1}^C \sum_{v \in S_c} \sum_{w \in S_c} (p(v,w)+p(w,v)-2p_V(v)p_V(w)).
\eeq
As $\tilde p(v,w)=(p(v,w)+p(w,v))/2$, we have
$$\tilde p_V(v)=\sum_{w \in V_g}\tilde p(v,w)=p_V(v).$$
 Thus,
$$
Q({\cal P})=\sum_{c=1}^C \sum_{v \in S_c} \sum_{w \in S_c} (\tilde p(v,w)-\tilde p_V(v)\tilde p_V(w))=\tilde Q ({\cal P}).
$$

\subsection*{Appendix D}

In this section, we prove \rthe{community}.

(i) Since we choose two sets that have a nonnegative correlation measure, i.e., $\qq (S_i, S_j)\ge 0$, to merge, it is easy to see from \req{random4400} and \req{random4407} that the modularity index is non-decreasing in every iteration.

(ii)
Suppose that there is only one set left. Then this set is $V_g$ and it is the trivial community.
On the other hand, suppose that there are $C \ge 2$ sets $\{S_1, S_2, \ldots, S_C\}$ left when the algorithm converges.
Then we know that $\qq(S_i, S_j)<0$ for $i \ne j$.

Note from \req{random4405} and  \req{random4406} that for any node $v$,
\beq{random4415}
\qq(\{v\}, V_g)=\sum_{w \in V_g}\qq(v, w)=0.
\eeq
Thus,
\beq{random4417}
\qq(S_i, V_g)=\sum_{v \in S_i}\qq(\{v\}, V_g)=0.
\eeq
Since $\{S_1, S_2, \ldots, S_C\}$  is a partition of $V_g$, it then follows that
\beq{random4418}
0=\qq(S_i, V_g)=\qq(S_i , S_i) + \sum_{j \ne i}\qq(S_i, S_j).
\eeq
Since $\qq(S_i, S_j)<0$ for $i \ne j$, we conclude that
$\qq(S_i,S_i)>0$ and thus $S_i$ is a community.

(iii) Suppose that $S_i$ and $S_j$ are merged into the new set $S_k$.
According to the update rules in the algorithm and the symmetric property of $\qq(\cdot, \cdot)$, we know that
\bearn
&&\qq (S_k, S_\ell) =\qq(S_\ell, S_k)=\qq(S_i, S_\ell)+\qq(S_j,S_\ell)\\
&&=\qq(S_i, S_\ell)+\qq(S_\ell, S_j),
\eearn
 for  all $\ell \ne k$.
Thus,
$$\bar \qq(S_k, S_\ell)={{|S_i|} \over {|S_i|+|S_j|}} \bar \qq(S_i, S_\ell)+
{{|S_j|} \over {|S_i|+|S_j|}} \bar \qq(S_\ell,S_j).$$
Since we select the two sets  with the {\em largest} average correlation measure in each merge operation,
we have $\bar \qq(S_i, S_\ell) \le \bar \qq (S_i, S_j)$ and
$\bar \qq(S_\ell,S_j) \le \bar \qq (S_i, S_j)$. These then lead to
$$\bar \qq(S_k, S_\ell) \le \bar \qq (S_i, S_j).$$
Thus, $\bar \qq (S_i, S_j)$ is not less than the average correlation measure between any two sets after the merge operation. As such, the average correlation measure at each merge is non-increasing.

\end{document}